\documentclass{emulateapj}

\usepackage{graphicx}
\usepackage{pjournal}

\newcommand{\bit}{\begin{itemize}}
\newcommand{\eit}{\end{itemize}}
\newcommand{\bce}{\begin{center}}
\newcommand{\ece}{\end{center}}

\newcommand{\bfr}{\begin{flushright}}
\newcommand{\efr}{\end{flushright}}
\newcommand{\ben}{\begin{enumerate}}
\newcommand{\een}{\end{enumerate}}
\newcommand{\bfi}{\begin{figure}}
\newcommand{\efi}{\end{figure}}
\newcommand{\beq}{\begin{equation}}
\newcommand{\eeq}{\end{equation}}

\newcommand{\ump}{$\mu$m}
\newcommand{\um}{$\mu$m }

\begin{document}

\title{IRIS : A new generation of IRAS maps}

\author{Marc-Antoine Miville-Desch\^enes}
\affil{Canadian Institute for Theoretical Astrophysics, 60 St-George st, 
Toronto, Ontario, M5S 3H8, Canada; mamd@cita.utoronto.ca}
\and
\author{Guilaine Lagache}
\affil{Institut d'Astrophysique Spatiale, Universit\'e Paris-Sud, B\^at. 121,
91405, Orsay, France; lagache@ias.u-psud.fr}

\begin{abstract}
The Infrared Astronomical Satellite (IRAS) had a tremendous impact on
many areas of modern astrophysics. In particular it revealed 
the ubiquity of infrared cirrus that are a spectacular manifestation
of the interstellar medium complexity but also an 
important foreground for observational cosmology. 
With the forthcoming Planck satellite there is a need for all-sky 
complementary data sets with arcminute resolution that can bring informations 
on specific foreground emissions that contaminate the Cosmic Microwave 
Background radiation. With its $\sim 4'$ resolution matching
perfectly the high-frequency bands of Planck, IRAS is a natural data set
to study the variations of dust properties at all scales.
But the latest version of the images delivered by the IRAS team
(the ISSA plates) suffer from calibration, zero level and striping problems
that can preclude its use, especially at 12 and 25~\ump.
In this paper we present how we proceeded to solve each of these problems 
and enhance significantly the general quality of the ISSA plates in the 
four bands (12, 25, 60 and 100~\ump).
This new generation of IRAS images, called IRIS, benefits from 
a better zodiacal light subtraction, from a calibration and zero level
compatible with DIRBE, and from a better destriping.
At 100~\um the IRIS product is also a significant improvement from 
the Schlegel et al. (1998) maps. IRIS keeps the full ISSA resolution, it 
includes well calibrated point sources and the diffuse emission calibration
at scales smaller than $1^\circ$ was corrected for the variation of the 
IRAS detector responsivity with scale and brightness.
The uncertainty on the IRIS calibration and zero level 
are dominated by the uncertainty on the DIRBE calibration and on the
accuracy of the zodiacal light model.

\end{abstract}


\section{Introduction}

From January to November 1983 the Infrared Astronomical Satellite (IRAS), 
a joint project of the US, UK and the Netherlands \cite[]{neugebauer84},
performed a survey of 98\% of the sky at four wavelengths :  
12, 25, 60 and 100 \ump. 
IRAS lead to numerous scientific discoveries spanning a broad range of 
astrophysical subjects, from comets, circumstellar disk to interacting 
galaxies. The satellite was designed to optimize the reliablility of point 
sources detection and photometry;
one of the great legacies of IRAS is certainly its catalog of 
more than 250~000 point sources. On the other hand, the relative stability 
of its detectors also allowed mapping
of extended emission. In fact IRAS made a significant contribution to our 
understanding of Galactic diffuse emission by revealing the interstellar dust 
emission of infrared cirrus, that can be observed in any direction on the 
sky \cite[]{low84}.

A first set of extended emission maps (the SkyFlux atlas) 
was released in 1984 and 1986 along with the IRAS Point Source Catalog.
It was then clear that significant improvements of the sensitivity 
and photometric accuracy could be obtained based on the acquired knowledge 
of the instrument. The IRAS data were later reprocessed based
on the knowledge of the instrumental response available at that time but also 
by applying a zodiacal light correction and a destriping method. 
This second generation processing, which increased
the sensitivity by a factor up to five, lead to
the the IRAS Sky Survey Atlas (ISSA) published in 1991 and 1992. 
The IRAS Sky Survey Atlas Explanatory Supplement \cite[]{wheelock93}
describes in detail the processing and analysis of this atlas.

Since its publication, the ISSA has been a tremendously useful data product,
used to study various aspects of Galactic (e.g. \cite{boulanger88})
and extra-galactic (e.g. \cite{miville-deschenes2002b}) 
diffuse emission. The ISSA plates
have become an essential data set for any multi-wavelength analysis of the 
interstellar medium.  
Unfortunately, even after the heroic processing done by the IRAS team, 
the ISSA plates still suffer from defects (striping, calibration, 
zero level, zodiacal light) that can limit significantly their use. 
In the context of the numerous ongoing and future 
infrared/submm/mm missions (Spitzer, Herschel, Planck, JWST, ASTRO-F...), 
IRAS data could still be highly valuable, 
even with their relatively coarse angular resolution ($\sim$4').
This was our main motivation
to reprocess the ISSA plates of all IRAS bands, in order to improve 
their sensitivity and absolute calibration. 

This new generation of IRAS maps, 
a product named IRIS (for Improved Reprocessing of the IRAS Survey), 
will be of particular interest for the Planck mission dedicated
to the study of the Cosmic Microwave Background (CMB). Planck will perform
an all-sky survey in the 350~\ump-10~mm wavelength range with an angular 
resolution in the high frequency bands comparable to IRAS. 
One of the main anticipated challenges of this mission 
 will be the
separation of all the components, including dust, that contribute 
to the signal in this wavelength range (as it is for WMAP, \citep[]{bennett2003a}).
Several studies of dust emission at high Galactic latitudes
showed large variations of dust properties at all scales,
even in the diffuse interstellar medium 
\cite[]{boulanger90,lagache98,bernard99,miville-deschenes2002,stepnik2003}.
To improve on the dust models available (e.g. \cite{finkbeiner99}) 
and bring them to an accuracy
level required for Planck, dust evolution should be understood 
in a more general way, including the properties of the 
smaller grain populations that give some informations on the 
evolutionary stage of dust (coagulation, fragmentation) and 
on the properties of the bigger grains that contribute to 
the submm/mm emission.
Small grains have also been suggested \cite[]{draine98} to
be responsible for the anomalous microwave emission 
(e.g. \cite{lagache2003b}).
In this context, well calibrated, arcminute resolution and
all-sky dust emission maps at 
12, 25, 60 and 100~\um would be of tremendous help to understand
and model dust emission in the Planck bands.
This is one of the main motivations of the present work.

The present paper describes in detail the procedure we
followed to reduce significantly the striping of the ISSA maps and to 
rescale them to have an absolute calibration
coherent with both the DIRBE and IRAS Point Source Catalog photometry.
In \S~\ref{section_issa}
we describe in detail the ISSA product. Our reprocessing 
of the ISSA plates, which include an automatic deglitching, 
an improved destriper and an absolute calibration and zero level correction, 
is described in \S~\ref{section_cleaning} and \S~\ref{section_calibration}. 
The overall properties of the IRIS product are discussed 
in \S~\ref{section_properties}.
The whole IRIS data set is available at {\bf http://www.cita.utoronto.ca/~mamd/IRIS}
and at {\bf http://www.ias.u-psud.fr/iris/}. 

\section{The IRAS Sky Survey Atlas}

\label{section_issa}

\subsection{Processing of the ISSA plates}

During its 10 months operation period, IRAS made two surveys of 98\% of the 
sky and a third one of 75\% of the sky. To confirm point source detection, 
each survey, called an HCON for Hours CONfirmation, 
consisted of two coverages of the sky separated by up to 36 hours. 
The first two HCONs (HCON-1 and HCON-2) were done concurrently with 
the second one behind the first one by a few weeks. The third survey (HCON-3) 
began after the first two were completed. Due to
exhaustion of the liquid helium supply the third HCON could not be completed 
and it thus covers only 75\% of the sky.

The ISSA is a set of 430 fields of $500\times500$ pixels of 1.5'. 
Each field is a $12.5^\circ \times 12.5^\circ$ region centered every 
$10^\circ$ in declination. All HCONs represent the sky surface brightness at 
the 4 wavelengths from which an estimate of the zodiacal light was subtracted. 
The HCONs were destripped, checked for zero level stability and visually 
examined for artifacts and glitches. The processed HCONs were then co-added, 
using sky coverage maps, to produce one final map (named HCON-0) per field 
and per wavelength.

\subsection{Diffuse emission photometry}

Special care has been taken to have a consistent diffuse emission 
calibration through the Atlas. But, as the IRAS mission was designed 
to provide absolute photometry for point sources only,
the ISSA images give only relative photometry on spatial scales
larger than $\sim$5' and cannot be used
directly to determine the absolute surface brigthness of diffuse emission.
The uncertainty on the zero level is dominated by
uncertainties in the zodiacal light model at 12 and 25 \ump, 
and by detector drifts at 60 and 100 \ump. 

As it was designed to study the cosmic infrared background, 
the DIRBE (Diffuse Infrared Background Experiment) experiment on board the 
COBE (Cosmic Background Explorer) satellite has made absolute surface 
brightness measurements of the sky at scales larger than 0.7$^\circ$. 
Moreover the DIRBE data were corrected with a better zodiacal light model 
\cite[]{kelsall98} than the one used for the ISSA.
Fortunately DIRBE and IRAS had very similar filters and therefore the DIRBE 
data can be used to determine the absolute IRAS calibration at scales larger 
than 0.7$^\circ$. Some efforts have been put in the comparison between DIRBE 
data and individual HCON images. \cite{wheelock93} showed that 
the amplitude of the 60 and 100 \um intensity fluctuations 
are overestimated in the ISSA maps with respect to DIRBE. 
\cite{wheelock93} suggest a linear relation between the IRAS and DIRBE 
surface brightness that also takes into account the difference in zero level:
\begin{equation}
I_{\lambda}^{DIRBE}(\alpha, \delta) = G_\lambda \times I_{\lambda}^{ISSA}(\alpha, \delta) 
+ B_{\lambda}(\alpha,\delta).
\end{equation}
The gain correction factors ($G_\lambda$) given in Table~\ref{table_issa} 
were originally obtained on spatial scales of the order of an IRAS scan 
($\sim 8^\circ$).
We have performed a detailed comparison of the DIRBE and IRAS data 
on numerous regions covering a broad range of brightness and 
we conclude that these factors vary with angular scale and brightness.
This issue is discussed in detail in \S~\ref{section_gain}.
Regarding the zero point of the ISSA maps, \cite{schlegel98} showed that the 
100 \um offset $B_{100}(\alpha,\delta)$ varies significantly across the sky.
Our analysis show that this is the case at all wavelengths 
(see \S~\ref{section_offset}).

\begin{table*}
\begin{center}
\caption{\label{table_issa} Compilation of IRIS related quantities}
\begin{tabular}{lcccc}\hline
 & \multicolumn{4}{c}{Wavelength (\ump)}\\ \cline{2-5}
Quantity & 12 & 25 & 60 & 100\\\hline
1. IRAS resolution (arcmin) &  $0.75\times 4.5$ &  $0.75\times 4.6$ &
$1.5\times 4.7$ &  $3.0\times 5.0$\\
2. IRIS resolution (arcmin) & $3.8\pm0.2$ & 
$3.8\pm0.2$ & $4.0\pm0.2$ & $4.3\pm0.2$ \\
3. ISSA noise level (MJy/sr) & $0.04\pm0.01$ & $0.05\pm0.02$ & $0.04\pm0.01$ & $0.07\pm0.03$ \\
4. IRIS noise level (MJy/sr) & $0.04\pm0.01$ & $0.05\pm0.02$ & $0.03\pm0.01$ & $0.06\pm0.02$ \\
5. DC/AC & 0.78 & 0.82 & 0.92 & 1.00\\
6. $G_\lambda$ & $1.06\pm0.02$ & $1.01\pm0.02$ & $0.87\pm0.05$ & $0.72\pm0.07$\\
7. DIRBE S(gain) (\%) & 5.1 & 15.1 & 10.4 & 13.5\\
8. DIRBE S(offset) (MJy/sr) & $6.0\times10^{-5}$ & $8.3\times10^{-5}$ & 
0.027 & 0.027 \\\hline
\end{tabular}
\end{center}
Note. -- Row 1: Effective instrumental resolution of the IRAS 
detectors \cite[]{wheelock93}. Row 2: Angular resolution of the ISSA and IRIS
plates (see \S~\ref{section_resolution}). 
Row 3-4: The ISSA and IRIS noise level given here are representative
of the median instrumental noise of the whole plates. 
The noise level varies from one plate to the other as a function of coverage
(see \S~\ref{section_noise}). Row 5-6: DC/AC and IRAS/DIRBE corrections given 
by the IRAS team \cite[]{wheelock93}.  
Row 7-8: DIRBE gain and offset uncertainties from \cite{hauser98}.
\end{table*}

\subsection{Point Source Photometry}

By staring at point sources, the IRAS team clearly showed 
\cite[]{wheelock93} that the instantaneous (AC) and long-term (DC) 
responsivity of the IRAS detectors were different. For a survey
mission like IRAS, this introduces a responsivity difference 
between point-like object and diffuse emission. The ISSA plates
are DC calibrated which means that point source fluxes estimated
on a ISSA plate have to be corrected for the DC/AC responsivity
difference (see Table~\ref{table_issa}).
Even if the ISSA product was not optimized for point sources
\cite{wheelock93} showed that, when corrected for the appropriate
DC/AC factor, point source fluxes measured
on ISSA plates are photometrically consistent with the IRAS Point Source 
Catalog to within 10\%.
We have verified this by comparing the DIRBE and ISSA (convolved by the 
DIRBE beam) fluxes of several stars. We also found a general good 
agreement between these two data sets. 

\subsection{Post-processing of the ISSA plates}

Beside the responsivity variation and zero level problems, 
the ISSA plates have also relatively strong stripes,
related to the scanning on the sky, that participate significantly 
to the noise. The main goals of the present work were first to reduce the
amplitude of those stripes to lower the noise level and second to
correct the ISSA plates calibration to match both the DIRBE 
and the Point Source Catalog photometry.

In the next sections we describe how we have addressed all of
these problems. First we discuss how we have deglitched
automatically the individual HCONs (\S~\ref{section_deglitching}). 
Then we describe our destriping algorithm (\S~\ref{section_destriping}). 
Finally we describe how we use the comparison of the IRAS and 
DIRBE data to establish a calibration and a zero level correction that is 
coherent both for point sources and diffuse emission 
(\S~\ref{section_calibration}).

Through out the processing we have taken great care to work
on individual HCONs to improve the destriping of the maps and 
to be able to estimate the noise level in the final averaged maps. 
This specific point is discussed in \S~\ref{section_noise}.

\section{Cleaning the ISSA plates}

\label{section_cleaning}

\subsection{Deglitching}

\label{section_deglitching}

Before co-adding the individual HCONs, the IRAS team inspected them 
visually 
to identify artifacts like moving objects (e.g. asteroids) or residual 
glitches. To allow for further validity checks of point sources seen 
(or not) in the co-added maps, the IRAS team decided to provide the 
individual HCONs that include those artifacts. As we are working directly 
on the individual HCONs we had to redo the deglitching but we took 
advantage of the fact that the co-added maps (HCON-0) have already been 
deglitched. 

First, using the method described in \S~\ref{section_point_source}
we identified point sources both in the individual HCONs and in the co-added
map. In a given map the identified pixels were then gathered together into
regions by consecutively labelling all the detecting regions
with a unique region index
\footnote{We use the label\_region function of IDL}.
Then we computed the central position 
(taken as the position of the brightest pixel) and the maximum radius
of each region. For each region found in the individual HCON (1,2 or 3) we 
derived the distance to all regions found in the co-added map.
Each HCON region with no counterpart in the co-added map 
at distance lower than the sum of the two regions maximum radius
is considered as a glitch.
The corresponding pixels in the HCON are then set to the background 
level at that position. 

This automatic method worked very well to identify automatically residual glitches 
and moving objects on most of the ISSA plates. It was not as performant in
very bright regions were the diffuse emission varies significantly
at small scales. Therefore we did not deglitch regions brighter that
200 MJy/sr at 100~\ump. We also dealt manually with Saturn due to its
large extend in the affected HCONs.

\subsection{Destriping}

\label{section_destriping}
\begin{figure}
\includegraphics[width=\linewidth]{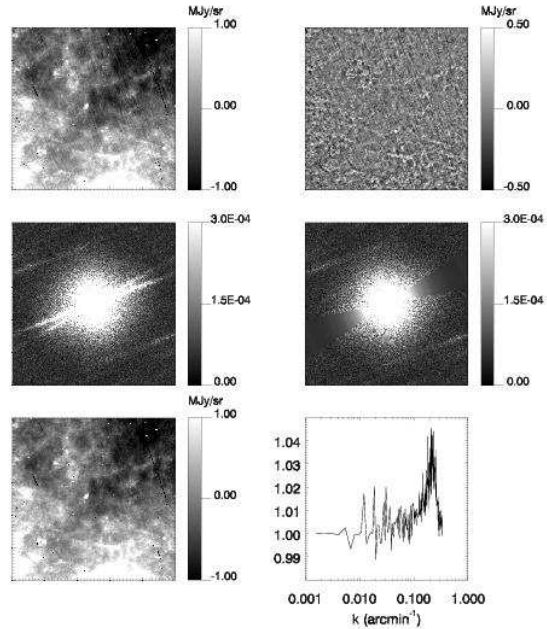}
\caption{\label{HCON_stripes} Illustration of the destriping algorithm.
The image (plate number 83 at 60 $\rm \mu$m) is choosen such that it 
contains point sources, diffuse emission and undefined regions. 
Top-left: HCON; Top-right: wavelet filtered HCON; Middle-left:
FFT of the filtered HCON; Middle-right: modified FFT; Bottom-left: destriped 
HCON; Bottom-right: ratio of the power spectrum of the HCON and the destriped 
HCON.}
\end{figure}

The ISSA plates have relatively strong stripes that are related to the 
scanning on the sky. We describe in this part our destriping algorithm that 
reduces significantly the amplitude of those stripes.
The destriping is done in Fourier space for each HCON. For typical HCONs,
the stripes have one or two directions but there exists more complicated 
regions where there are several directions (up to 5 for example near the 
Large Magellanic Cloud). In the Fourier space, such stripes give extra-power 
in well-defined azimuthal directions that can be identified and corrected 
(Fig. \ref{HCON_stripes}).

To be able to identify clearly the striping pattern in Fourier space, two
effects have to be dealt with: point-like sources and large scale gradients.
Point sources introduce contaminating power in Fourier space
in every direction and with a flat power spectrum. Their main effect is to
lower significantly the contrast of stripes in Fourier space, especially
at large wavenumbers $k$. On the contrary,
a large scale gradient across an image produces extra-power 
in privileged directions in Fourier space (horizontal and vertical)
that mimics stripes.
To get rid of this effect it is quite common to apodize the image (see \cite{schlegel98} 
for example) but this destroies the information on the edge of the image. 

To get round these difficulties we used a wavelet decomposition 
(``a trou'' algorithm) of every HCON and worked 
only on an image that is the sum of the first five scales 
(2, 4, 8, 16 and 32 pixels of 1.5'). This filtered map contains only small scale
structures, usually dominated by the stripes, and is thus free of any 
large scale gradient (Fig. \ref{HCON_stripes}). 
To increase the contrast of the striping pattern we removed point sources
at each scale of the wavelet decomposition prior to summing them; 
we cliped all the wavelet coefficients higher than 7$\sigma$, where $\sigma$ 
is computed at each scale. 
The original HCON was then decomposed into
three components: a large scale emission map (scales larger than 32 pixels),
small scale emission map (scales smaller than 32 pixels) and a point sources 
map. The following destriping algorithm was only applied on the small scale
emission map.

Each small scale emission map was Fast Fourier Transformed (FFT) and
the averaged magnitude of the FFT was computed as a function of 
angle $\theta$ in 360 bins of $1^\circ$ wide. 
We identified the bins with averaged magnitude greater than N$\sigma$ 
and replaced them by the averaged azimuthal magnitude.
This is equivalent to replace the identified bins by the power spectrum
of the clean bins. This procedure was repeated twice 
on artificially rotated FFT pattern (to detect stripes with direction falling between two bins). 
During this process, we progressively decreased N from 7 to 5. 
We then made the inverse transform of the cleaned FFT and added the result
to the large scale emission and point source maps.

This procedure can only be applied on fully defined HCONs. 
For each plate, the missing data were iteratively replaced by the destriped 
data of the companion HCONs. When large regions have not been observed by the 
3 HCONs, we used the DIRBE data, corrected for the appropriate gain and 
offset values.

\begin{figure*}
\includegraphics[width=\linewidth]{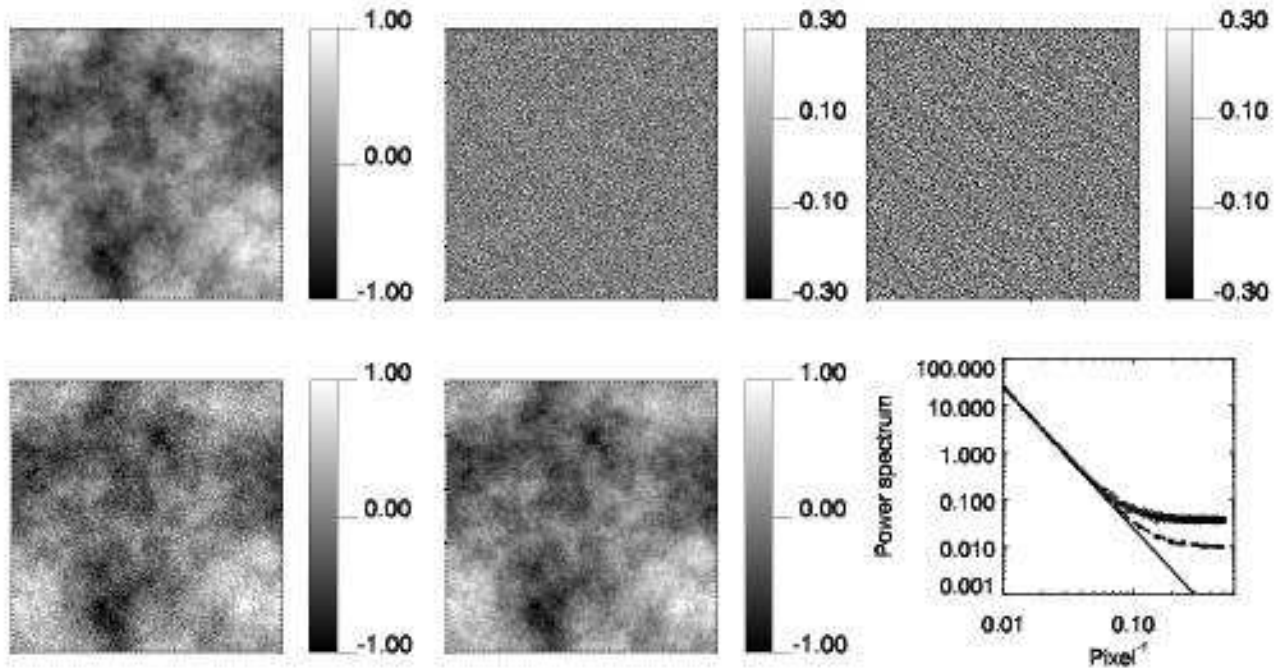}
\caption{\label{simu_destripe} Simulated high latitude cirrus region with 
noise and stripes. Top-left: cirrus (P(k)$\propto$ k$^{-3}$); 
Top-middle: random noise; Top-right: stripes;
Bottom-left: total simulated map; Bottom-middle: destriped map;
Bottom-right: power spectrum of the cirrus (continuous line),
cirrus+noise (dashed line), cirrus+noise+stripes (diamonds),
and the recovered map after the destriping process 
(dashed-dotted line), in arbitrary units. The dashed and 
dashed-dotted lines are superimposed (the difference is at the order
of 3$\%$ at very small scales).
}
\end{figure*}

Our procedure has been extensively tested on simulations.
We simulated typical interstellar diffuse regions using 
fractional Brownian motion images
\cite[]{miville-deschenes2003b} with a similar power spectrum to 
what is observed in the ISM (spectral index of -3 \cite[]{gautier92}). 
We added to this map typical random noise together with stripes that were
simulated in Fourier Space (note that we cannot inject in our simulations
empirically derived stripes from real data since such maps, obtained
by wavelet filtering real data visually dominated by stripes, are also
contaminated by the instrumental noise). 
The simulated map were then processed through the whole 
destriping procedure.
An example of such a simulated map  for a typical high latitude region
at 12~$\rm \mu$m (where the amplitude of the stripes are of the same 
order than the signal) is shown
on Fig. \ref{simu_destripe}, together with the result of the destriping algorithm.
We computed the power spectrum of the input noisy interstellar map and that 
of the recovered map. In such regions, we modified the power spectrum at small
scales by only about 3$\%$. This demonstrates that we have supressed 
the contribution of the stripes (that was four times higher than the signal 
on the power spectrum at small scales in this case) without adding extra-noise. In 
brighter regions, the contribution of the stripes becomes smaller compared 
to the galactic emission, leading to a still better recoverage 
of the simulated map. These simulations clearly show the robustness 
and efficiency of our destriping process.

\section{Calibrating the ISSA plates}

\label{section_calibration}

\subsection{Infrared photo-detectors}

\label{sec_isocam}

\begin{figure}
\includegraphics[width=\linewidth]{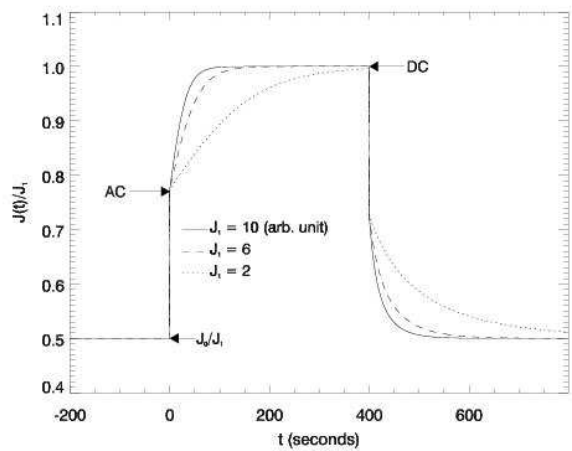}
\caption{\label{exemple_fouks} Illustration of the response of an
 infrared photo detector based on Eq.~\ref{fouks} \cite[]{fouks95}.
The response to upward and downward flux steps are shown for three
values of the flux $J_1$ = 2, 6 and 10. 
To illustrate the behavior of the detector when the flux value is double,
the lower flux value is $J_1/2$ for each case.
The response of the detector is very sensitive to the flux value; 
for lower flux value, the detector takes more time to stabilize.}
\end{figure}

\cite{wheelock93} showed that the responsivity of all IRAS detectors are 
affected by memory effects that introduce a calibration difference between
diffuse emission and point sources.
Since IRAS, and especially with the ISO mission, several
studies have been done to understand the response of infrared photo 
conductors like the IRAS 60 and 100 \um Ge:Ga detectors.
Using solid state and semiconductor physics,
\cite{vinokurov91} and \cite{fouks95} have studied the response of 
such detectors. They showed that the instantaneous response depends
on the illumination and on the history of the detector. A simplify
version of their model was successfully applied to ISOCAM observations
\cite[]{coulais2000}. This non-linear model is rather complex and it is 
not the purpose of the present paper to describe it in detail. But to
feel how the response of infrared photo detectors behaves, lets
consider the simple case of a detector stabilized at flux
$J_0$ that is suddenly illuminated at flux $J_1$ at time $t=0$.
In this case, the response $J(t)$ of the detector as a function of
time is described by:
\beq
\label{fouks}
J(t) = \beta J_1 + \frac{(1-\beta)J_1
  J_0}{J_0 + (J_1 - J_0) \exp (-t  J_1 / \lambda)}.
\eeq
The $\beta$ and $\lambda$ parameters depends on the physical
properties of the detector.
An example of this model is shown in Fig.~\ref{exemple_fouks}
where we illustrate what would be the response of a detector 
when the flux is double ($J_1 = 2 \times J_0$), for three values
of the initial flux $J_0$.
We see that the response of the detector is very sensitive 
to the flux value; for lower flux value, the detector 
takes more time to stabilize. Therefore, for a given 
integration time, an intensity fluctuation on a low 
surface brigthness region will be systematically 
underestimated in comparison to a brighter one.

This has an important consequence on survey
observations like the IRAS mission that scanned the sky
with a fixed integration time.
It is likely that, at low flux value,  
the integration time was shorter than the stabilization time of the
detector. Therefore the amplitude of the fluctuations on the sky 
are systematically underestimated in faint regions. 
This type of memory effect should leave another signature
on maps with a fixed integration time. 
As the satellite scans the sky it observes structures with various
angular scales. These structures can be seen as flux steps like
in Fig.~\ref{exemple_fouks}. Then, when the satellite goes through
a small scale structure, it is likely that the observed flux will never
reach the stabilization level. On the other hand, for very large scale
structures, the detector has enough time to stabilize. 
Therefore, the amplitude of the intensity fluctuations are
underestimated at low flux but also at small scales.

\subsection{Responsivity of the 12 and 25 \um detectors}

Similar effects were indeed observed by the IRAS team.
By staring at several stars of different brightness, 
\cite{wheelock93} showed that the responsivity $R$ 
of the 12 and 25 micron detectors depends on time and therefore on angular
scale, as the satellite is scanning the sky at a fixed rate
(3.85 arcmin/sec). The responsivity was observed to be constant 
for angular scales larger than 2$^\circ$ but it decreased
at smaller scales, in accordance with the model described earlier.
On the other hand the IRAS team reports no variations of
the responsivity as a function of brightness for these detectors.
In Table~\ref{table_gain} we give the responsivity
variation with angular scale from \cite{wheelock93} for the 12 and 
25~\um detectors but one must be aware that this variation of the 
responsivity with scale was not applied
to the ISSA plates. Instead a constant responsivity factor was used
to match the responsivity at large scale (i.e. the ISSA plates are 
DC calibrated) which means that the brightness fluctuations at scales
smaller than $\sim2^\circ$ are systematically underestimated at 12
and 25~\ump.

\begin{table*}
\begin{center}
\caption{\label{table_gain} Responsivity factors}
\begin{tabular}{ccc|ccccccc}\hline
$l$ & 12 \um & 25 \um & $l$ & & 60 \um & & & 100 \um\\
(degree) & $R_0$ & $R_0$ & (degree) &
$R_0$ & $A$ & $B$ & 
$R_0$ & $A$ & $B$ \\\hline
0.067 & 0.78 & 0.82 & 0.067 & 0.92 & 0.0 & 0.0 & 1.0 & 0.0 & 0.0\\
0.1 & 0.90 & 0.90 & 1.25 & 1.00 & 0.08 & -5.0e-5 & 0.96 & 0.008 & -4.8e-5\\
0.2 & 0.92 & 0.93 & 2.5 & 1.01 & 0.01 & -4.0e-5 & 0.99 & 0.011 & -6.5e-5\\
0.5 & 0.98 & 0.97 & 5.0 & 1.05 & 0.014 & -6.0e-5 & 1.08 & 0.012 & -8.0e-5\\
1.0 & 1.00 & 0.98 & 10.0 & 1.13 & 0.014 & -9.0e-5 & 1.25 & 0.08 & -5.9e-5\\
1.5 & 1.00 & 1.00 \\
2.0 & 1.00 & 1.00 \\\hline
\end{tabular}
\end{center}
The responsivity $R$ at a given wavelength, scale and brightness
is given by Eq.~\ref{eq_gain}. The responsivity at 12 and 25~\um does not
depend on brightness ($A=0$, $B=0$). The value of the responsivity at
the smallest scale (0.067 degree) is the DC/AC factor 
(see Table~\ref{table_issa}) needed to recover the correct point source
calibration.
\end{table*}

\subsection{Responsivity of the 60 and 100~\um detectors}

\label{section_gain}

At 60 and 100~\um the situation is complicated by the fact that the 
responsivity depends both on time (i.e. angular scale) and brightness.
\cite{wheelock93} did not give any prescription to address this
problem. To evaluate the variation of the responsivity with scale
and brightness of the 60 and 100~\um detectors 
we compared the IRAS and DIRBE data using a wavelet analysis.

To study the variations of the 60 and 100~\um responsivity
we have looked at variations of $\Delta IRAS/\Delta DIRBE$ as 
a function of scale and brightness.
For a given diffuse structure on the sky the quantity 
$\Delta IRAS/\Delta DIRBE$ 
represents the ratio of the brightness fluctuation (above the background) of 
that structure seen in the IRAS and DIRBE data. If we take the DIRBE 
calibration as an absolute reference, the quantity $\Delta IRAS/\Delta DIRBE$ 
is equal to the IRAS responsivity $R$ to a brightness fluctuation.
To compute  $\Delta IRAS/\Delta DIRBE$ as a function of scale and brightness 
we first put the two dataset on a common
grid and at the same angular resolution.
For each ISSA plate we have constructed a larger map ($600\times600$ pixels)
by mosaicking neighbouring plates. This step is necessary in order to
avoid edge effects in further convolutions. Next this bigger ISSA plate
is convolved by the DIRBE beam.
The DIRBE data are then projected on the ISSA grid.
Finally both maps are convolved by a Gaussian beam of FHWM 40'
to smooth out the DIRBE pixelisation \cite[]{schlegel98}
and the extra 50 pixels wide edges are removed.
At that point the angular resolution and size of both maps are
of 1 degree and $500\times500$ respectively.

\begin{figure}
\begin{center}
\includegraphics[width=\linewidth]{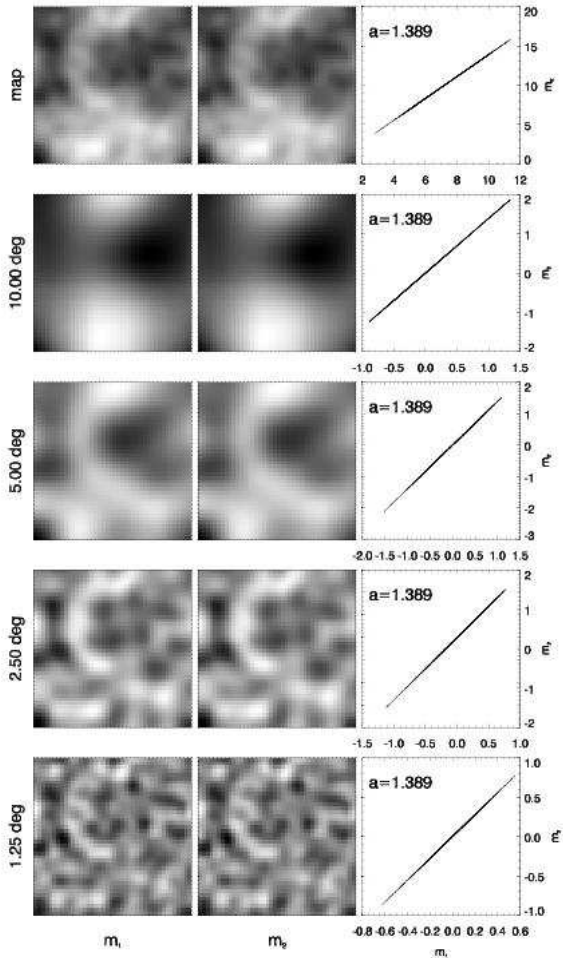}
\caption{\label{simul_wavelet}
Illustration of the use of the wavelet decomposition
to compute a gain difference between two maps at different scales.
Here the $m_1$ and $m_2$ are identical fBm images 
(shown in the top row) with a constant multiplicative factor 
between the two ($m_2 = 1.389 \times m_1$). The four
bottom rows shows the wavelet decomposition of $m_1$ and $m_2$.
The right column shows the correlation between $m_2$ and $m_1$
for the whole map and at each scale. The slope $a$
of the linear regression is given in each panel.}
\end{center}
\end{figure}

To compute the variation of $\Delta IRAS/\Delta DIRBE$ with scale and
brightness we used a wavelet transform (the ``a trou'' algorithm - 
\cite{starck98})
that allows us to study the brightness variations as a function of scale.
An exemple of such a decomposition is shown in Fig.~\ref{simul_wavelet}
on a fractional Brownian motion (fBm) image
\cite[]{miville-deschenes2003b} with a similar power spectrum to 
what is observed in the ISM (spectral index of -3) and convolved the
same way the IRAS and DIRBE data were.
On the top row of Fig.~\ref{simul_wavelet} we show two fBm images, 
$m_1$ and $m_2$,
where $m_2$ was constructed by simply applying a constant multiplicative
factor to $m_1$ ($m_2 = Q\times m_1$).
On the four bottom rows we show the wavelet coefficients at four differents 
scales (1.25, 2.5, 5 and 10 degrees). The important result
here is that the slope of the linear regression of the wavelet coefficients
gives exactly the multiplicative factor $Q$ at each scale.

\begin{figure}
\begin{center}
\includegraphics[width=\linewidth]{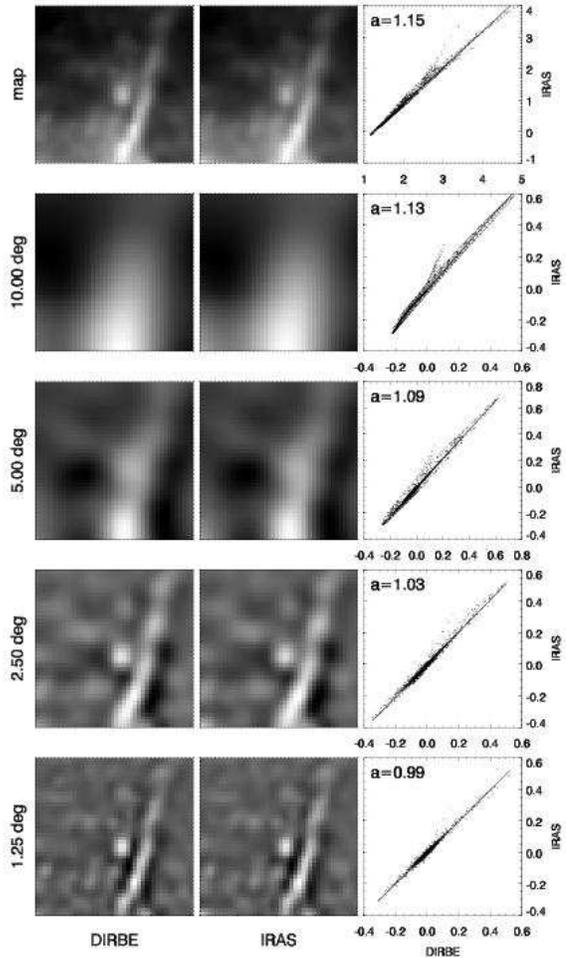}
\caption{\label{exemple_wavelet}
Example of a wavelet decomposition of a typical 60~\um image.
See Fig.~\ref{simul_wavelet} for details.}
\end{center}
\end{figure}

An exemple of the same wavelet decomposition
for a typical 60~\um field is shown in Fig.~\ref{exemple_wavelet}.
One would notice that the wavelet coefficients of the IRAS and DIRBE are
very well fitted by a linear fit with relatively small dispersion.
One noteworthy result here is the increase with scale of the slope 
($\Delta IRAS/\Delta DIRBE$) of the linear regression between
DIRBE and IRAS data. The trend seen 
in Fig.~\ref{exemple_wavelet} is totally
representative of what is generally observed.

\begin{figure}
\begin{center}
\includegraphics[width=\linewidth]{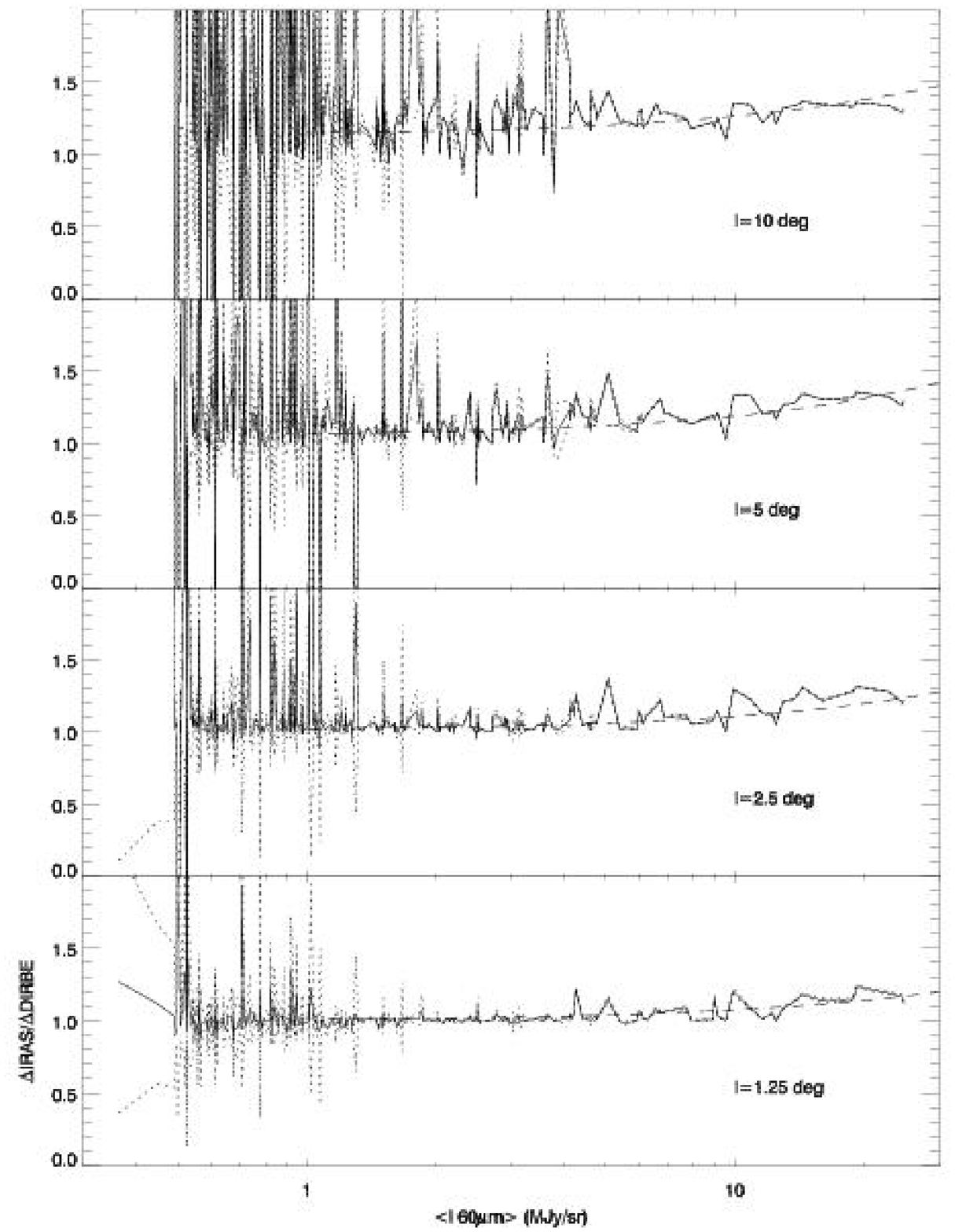}
\caption{\label{reponse_60} 
Responsivity of the 60~\um IRAS data compared to DIRBE as a 
function of the 60~\um DIRBE brightness and for four different scales.
The dotted lines are the slope of the linear regression of IRAS vs DIRBE
and DIRBE vs IRAS. The solid line is the average of the two dotted lines. 
The dashed line is the fit to the average slope using Eq.~\ref{eq_gain}.}
\end{center}
\end{figure}

\begin{figure}
\begin{center}
\includegraphics[width=\linewidth]{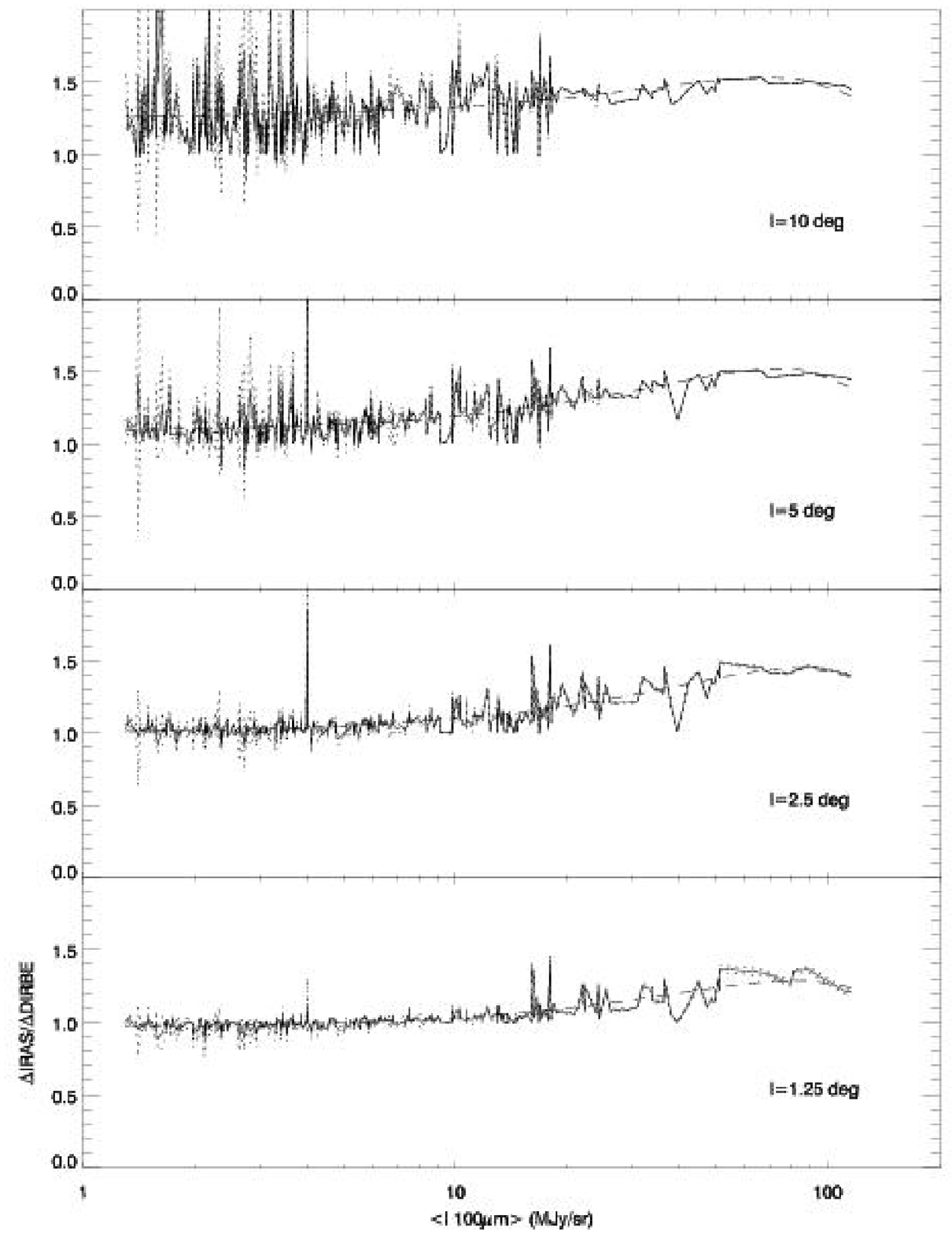}
\caption{\label{reponse_100} 
Responsivity of the 100~\um IRAS data compared to DIRBE as a function of 
the 100~\um DIRBE brightness and for four different scales. 
See Fig.~\ref{reponse_60} for details.}
\end{center}
\end{figure}

We have perform this analysis at 60 and 100~\ump, on all the ISSA plates 
with no undefined values, to also study the variation 
of $R=\Delta IRAS/\Delta DIRBE$
with brightness $<I_\lambda>${\footnote{Note here that, as stated at the
end of this section,  $<I_\lambda>$ is the median value
of the DIRBE data to limit the effect of variations of
the IRAS zero level}}. 
The results of this analysis are compiled in 
Figures~\ref{reponse_60} and \ref{reponse_100} where we show the variation
of $R$ with brightness at the four scales studied.
>From these two figures we found that the IRAS responsivity $R$ increases
with brightness for the two bands. The increase of $R$
with scale described previously is also seen in these two figures.
In addition the values of $R$ are systematically above 1.0
at large scale which confirms that IRAS brightness fluctuations are 
overestimated at large scales in the ISSA plates. The values obtained here at 
scales of 10$^\circ$ and for brightness of $\sim 3$ and $\sim 10$ MJy/sr at 
60 and 100~\um are compatible with the IRAS-DIRBE correction factors given 
by \cite{wheelock93} (see Table~\ref{table_issa}).

The increase of $R$ with scale and brightness indicates 
that the IRAS responsivity is generally lower for small scale
and faint brightness fluctuations than for large scale and
large brightness fluctuations.
This behavior is in complete accordance with what
is expected for typical photo-conductors (see \S~\ref{sec_isocam}).

The dotted lines in Figures~\ref{reponse_60} and \ref{reponse_100} 
represent
the slope of the linear regression of IRAS vs DIRBE and DIRBE vs IRAS, 
and the solid line is the average of the two. Computing both linear 
regressions allows us to conclude
that the trend observed here is not due to noise. The effect of 
detector noise is clearly seen
at small scale and at low brightness where the dispersion of the 
slope values is increasing.
There is also an increase of the dispersion at large scales 
(very clear at 60~\ump), due to the large scale drift 
of the detector responsivity or to a bad zodiacal light correction. 
These two effects produce an additive effect with spatial variations
at scales larger than a few degrees, affecting predominantly 
the slope measured at larger scales and at low brightness.

The curves in figures~\ref{reponse_60} and \ref{reponse_100} can be 
resonably well fitted by the following fonctional:
\begin{equation}
\label{eq_gain}
R = R_0 + A<I_\lambda>+ B<I_\lambda>^2.
\end{equation}
The values of $R_0$, $A$ and $B$, which all depend on
$\lambda$ and scale $l$, are given in Table~\ref{table_gain}.
The brightness $<I_\lambda>$ is the median value
of the DIRBE data, to limit the effect of variations of
the IRAS zero level.

\begin{figure}
\includegraphics[width=\linewidth]{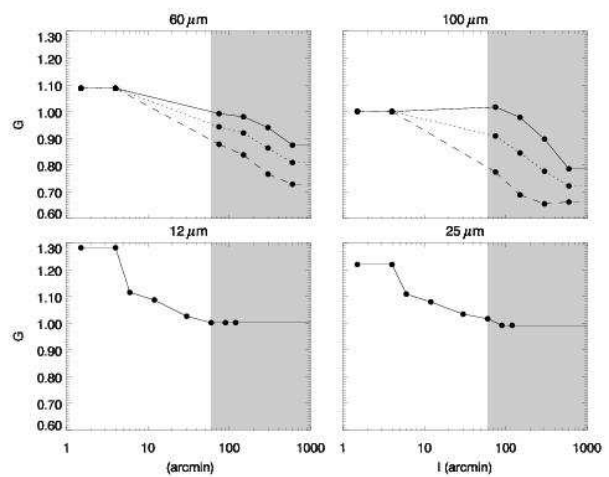}
\caption{\label{fig_gain} 
Responsivity correction factor ($G=1/R$) for the four bands as a function 
of angular scale. For the 60 and 100~\um bands the gain factor is given for 
three values of mean brightness. The solid, dotted and dashed lines are 
respectively for 1, 10 and 50 
MJy/sr at 100~\um and 0.3, 3 and 15 MJy/sr at 60~\ump. The shaded grey area 
represents the angular scales on which DIRBE data are used to scale the IRAS 
data.}
\end{figure}

\subsection{Reponsivity correction}

In the previous sections we showed that the IRAS responsivity to diffuse
emission varies with scale and, at 60 and 100~\ump, 
also with brightness. At these wavelengths we were able
to establish an analytical form of the responsivity variations 
(Eq.~\ref{eq_gain}) by comparing the IRAS and DIRBE data. 
This expression is also valid for the 12 and 25~\um bands; in this case 
it reduces to $R=R_0$ as the responsivity does not depend on the brightness
($A=0$ and $B=0$).

The responsivity correction to apply to the ISSA plates is the inverse of 
the expression given in Eq.~\ref{eq_gain} ($G=1/R$). 
In Fig.~\ref{fig_gain} we show the responsivity correction factor as function
of scale for the four bands. For the 60 and 100~\um bands we
present three curves to show the variation of the responsivity correction
with brightness.

\begin{figure}
\includegraphics[width=\linewidth]{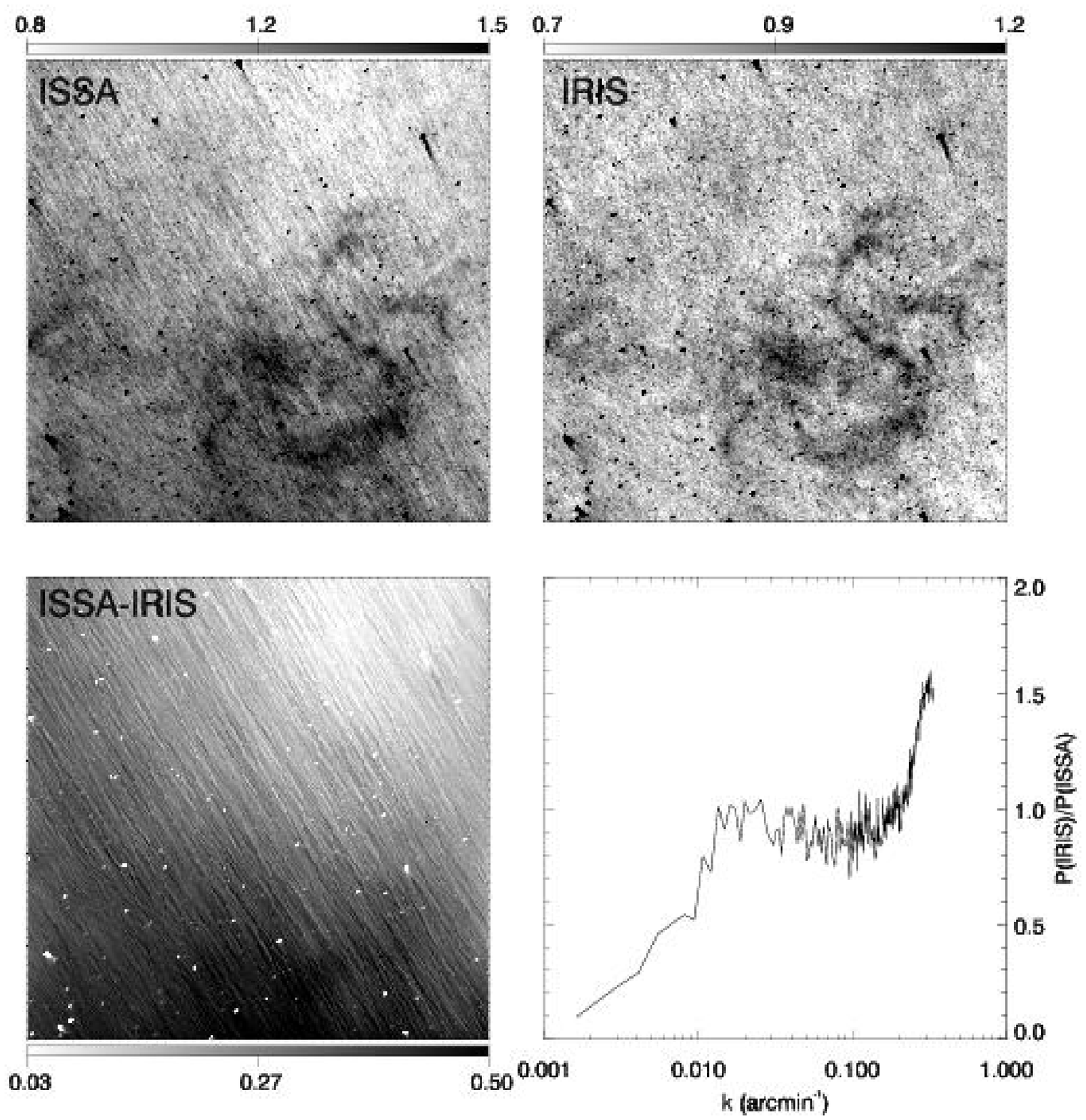}
\caption{\label{fig_avant_apres_12} Plate number 81 at 12~\ump.
Comparison of the ISSA (top-left) and IRIS (top-right) maps.
The difference between the two (ISSA-IRIS) is shown in the bottom-left
panel. The ratio of the power spectra of the IRIS and ISSA maps
is shown in the bottom-right panel. }
\end{figure}

\begin{figure}
\includegraphics[width=\linewidth]{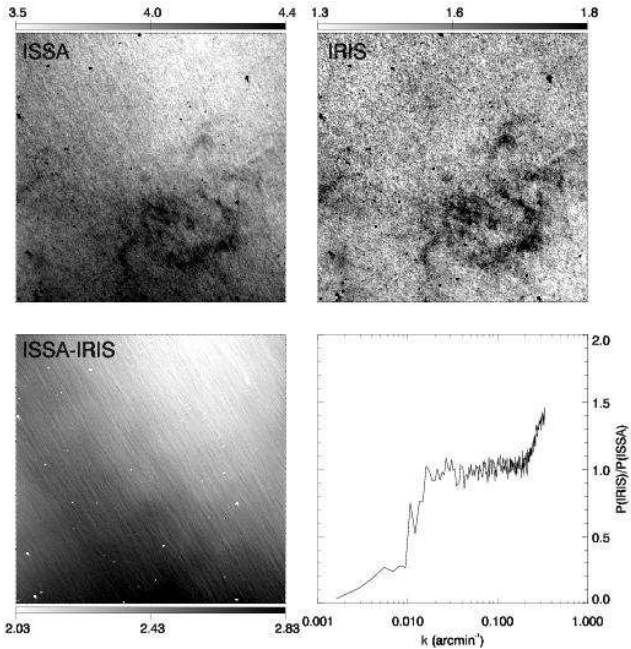}
\caption{\label{fig_avant_apres_25} Plate number 81 at 25~\ump.
See Fig.~\ref{fig_avant_apres_12}.}
\end{figure}

\begin{figure}
\includegraphics[width=\linewidth]{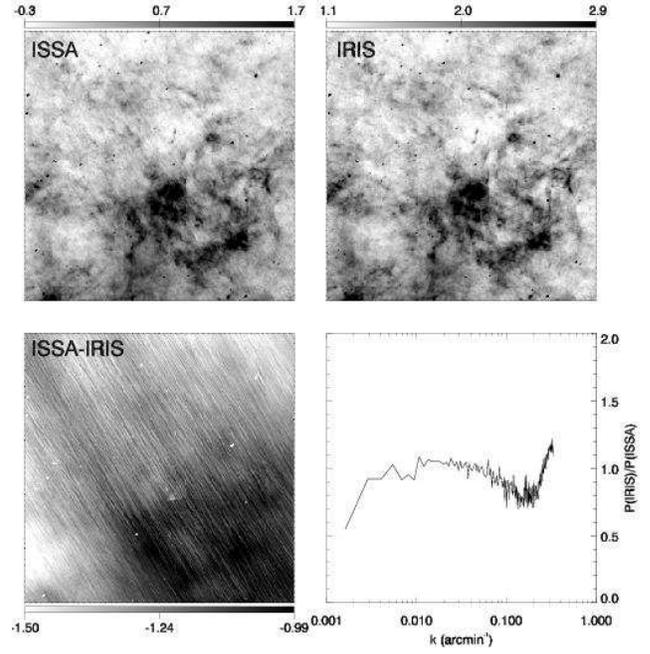}
\caption{\label{fig_avant_apres_60} Plate number 81 at 60~\ump.
See Fig.~\ref{fig_avant_apres_12}.} 
\end{figure}

\begin{figure}
\includegraphics[width=\linewidth]{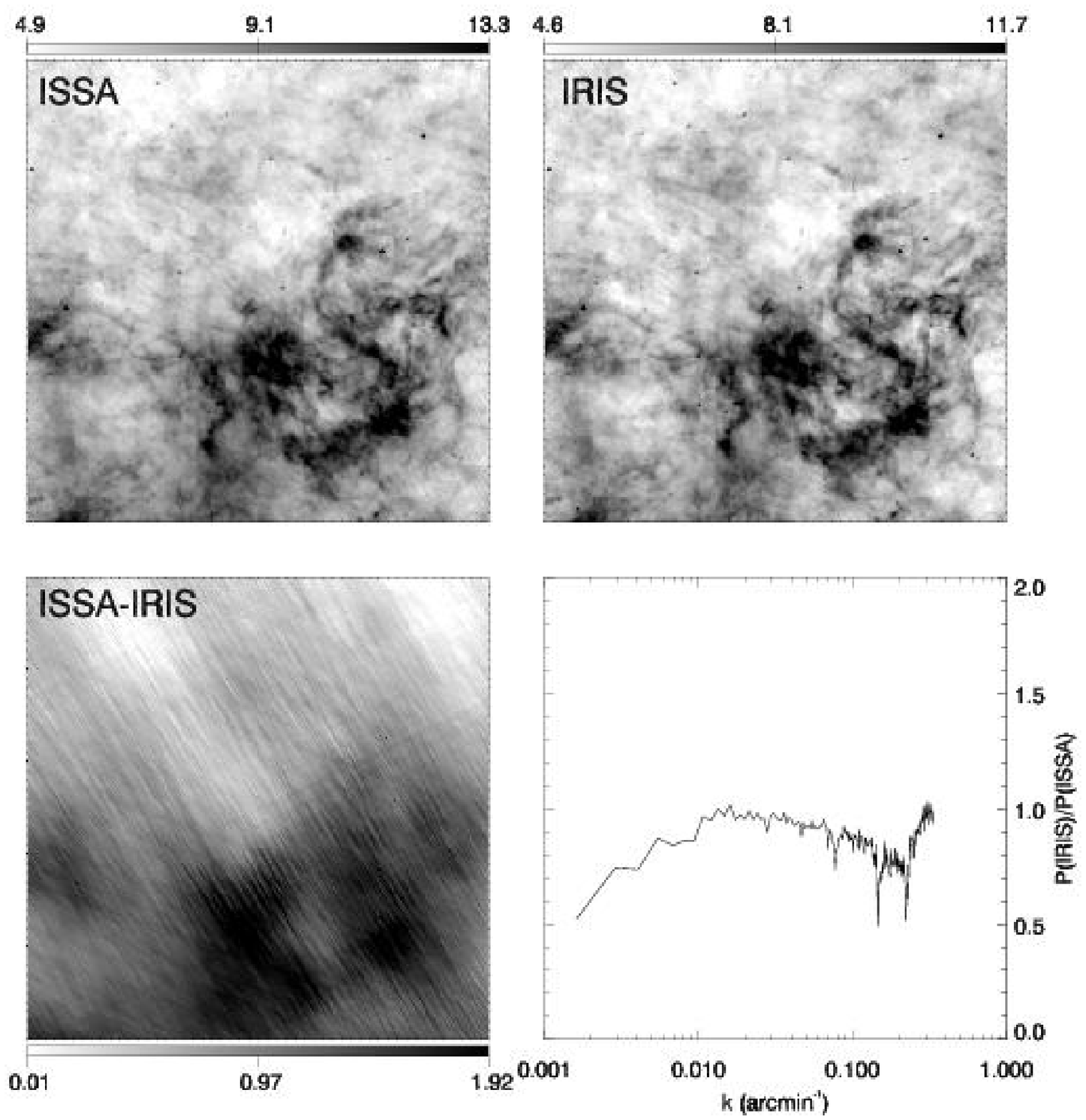}
\caption{\label{fig_avant_apres_100} Plate number 81 at 100~\ump.
See Fig.~\ref{fig_avant_apres_12}.}
\end{figure}

The implementation of such a correction is made difficult by the
fact that it depends on brightness (at 60 and 100~\ump). 
We need to apply a correction that depend on scale but
also on position to take into account brightness variations across
the map. To achieve this we use a local approch based on a 
wavelet decomposition of each ISSA HCON. 

To avoid discontinuity problems between adjacent plates 
each HCON is put in a $600\times600$ map and the undefined 
regions are filled using other HCONs or, when IRAS data are unavailable,
by DIRBE (properly scaled to take into account the gain difference
between the two data sets).
For each HCON at each wavelength, we first separated point sources from diffuse emission 
(see \S~\ref{section_point_source}).
To correct the diffuse emission map we decomposed it 
(using the ``a trou'' algorithm) for the first 6 scales 
(corresponding to 3', 6', 12', 24', 48', 1.6$^\circ$, 3.2$^\circ$).
Then we computed the proper gain correction factor at each scale 
and for each pixel of the map using Eq.~\ref{eq_gain}.
To compute this correction at 12 and 25~\um bands we simply use the 
responsivity values given in Table~\ref{table_gain} and interpolate 
at the 6 corresponding scales (from 3' to 3.2$^\circ$). At the 60
and 100~\um we do the same but we also use the DIRBE brightness
to compute the brightness dependance part of the 
responsivity correction.

The wavelet coefficients of the diffuse emission map were then
recombined to form the responsivity corrected diffuse emission map. Finally the
point sources map, properly calibrated by applying the corresponding
DC/AC factor (see Table~\ref{table_issa})
was added to the corrected diffuse emission map. 
The responsivity correction applied here does not correct the emission
at scales larger than 3.2$^\circ$ degrees. We take care of this in the next 
and step of the processing.

\subsection{Zero level correction}

\label{section_offset}

As it was mentioned earlier the zero level of the IRAS data 
is not well determined, due to detector drifts and/or residual
zodiacal emission. To correct this large scale
responsivity and the spatially varying zero level 
of the ISSA maps we again used the DIRBE data that were better 
calibrated and that used a better zodiacal emission model. Here we make the
assumption that the zero level does not vary on scales
smaller than 1$^\circ$ which is reasonable as detector drifts
and zodiacal emission are not expected to vary on such scales.

As for the responsivity correction we computed the offset map 
for every HCON, 
enlarged to $600\times600$ to make sure that we keep
the continuity between adjacent ISSA plates. Empty pixels were 
filled with the average map (HCON0), previously corrected
for the responsivity variation, or DIRBE data when necessary.
The HCON map was then convolved by the DIRBE beam and by a 
Gaussian beam (FWHM=40') to smooth out the DIRBE pixelisation. 
The DIRBE data were projected on this enlarged HCON map and convolved 
by the same Gaussian beam. At that point both data sets are on the same 
grid and at a $1^\circ$ resolution\footnote{It is worth mentioning that 
point sources were removed 
in the IRAS and DIRBE data to compute the offset. 
Even after the gain correction, which reconciliate the calibration
of point sources and diffuse emission, we had to remove point sources
for the offset computation as bright sources at 12 and 25~\um
have relatively strong tails that increase significantly their
flux when convolved by the DIRBE beam.}.
We then computed the difference
between the two maps (DIRBE-IRAS). This difference map was added
to the HCON. When all the 3 HCONs were corrected we computed the
average map (HCON-0) using the appropriate coverage maps.
This is the final step of the processing.

\begin{figure}
\includegraphics[width=\linewidth]{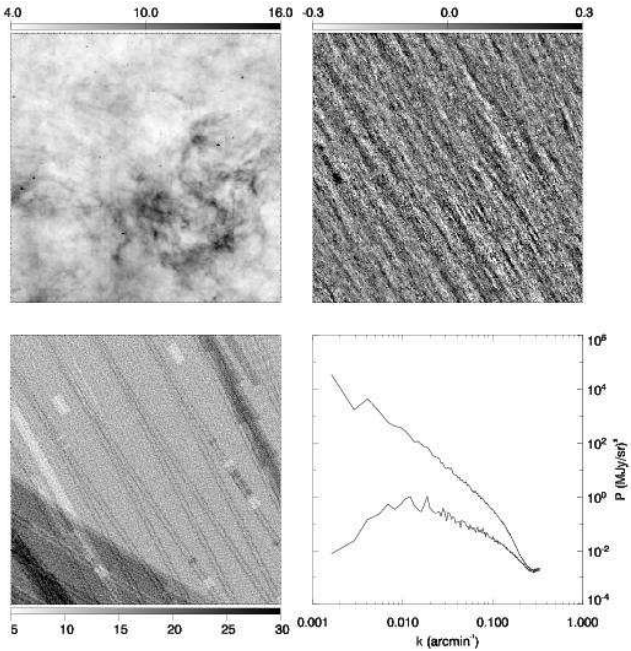}
\caption{\label{fig_noise_iris} {\bf Top:} 
Typical 100~\um IRIS image (left) 
together with its noise map (right). These two maps
are in MJy/sr. {\bf Bottom:} Coverage map (left) 
showing the number of data points averaged at each position ($N_{tot}(x,y)$). 
The power spectra of the 100~\um map and of its noise map
are shown in the right panel.}
\end{figure}

\section{Properties of the IRIS plates}

\label{section_properties}

The processing of the 4893 HCON plates was done using the McKenzie cluster
of the Canadian Institute for Theoretical Astrophysics. This Beowulf
cluster is composed of 512 2.4 GHz Xeon processors with a total of 256 GB of
RAM. We effectively used only 40 CPUs of the cluster at the same time
and the total computation time for the whole processing 
(deglitching, destriping, responsivity and offset correction) was 
$\sim$20 hours.\\

Figures~\ref{fig_avant_apres_12} to \ref{fig_avant_apres_100}.
present the ISSA and IRIS plate number 81
at the four wavelengths. Are also shown the difference between
the ISSA and IRIS maps and the ratio of their power spectra.
These figures show that the destriping described in \S~\ref{section_destriping}
removed very efficiently the stripes at all wavelengths. 
The second more noticeable improvement is the modification of
large scale structures in each maps, due to the responsivity and zero level
corrections. These two corrections removed very well residual
zodiacal light and detector drifts and improved significantly the
calibration of the data. The correction of the large scale structures
of the maps (that is due both to responsivity and zero level corrections)
is clearly seen in the ratio of the power spectra.

One interesting property seen on Figures~\ref{fig_avant_apres_12}
and \ref{fig_avant_apres_25} is that, even after 
the destriping, the power spectra ratio of the 12 and 25~\um images 
is $\sim 1$ on scales between 5' and 1$^\circ$. This is due to the 
responsivity correction applied (see Fig.~\ref{fig_gain}) that is higher 
than 1.0 at those scales. The noise reduction resulting in improved signal-to-noise
ratios obtained from the 
destriping is almost counterbalanced by the multiplication of all the 
small scale fluctuations (including noise) 
by a factor greater than 1. Nevertheless the noise properties 
of the IRIS plates are much more easy to handle as they are now
homogeneous.
On the other hand, at 60 and 100~\um the responsivity correction 
is not as important and there is a noise reduction that
is about 15-20$\%$ at scales 5-15'. \\

Finally, notice that the areas of the sky not covered by the 
IRAS survey are left empty in the IRIS product.

\subsection{Noise}

\label{section_noise}

Following the method described by \cite{miville-deschenes2002b}
one can estimate the power spectrum of the noise of each IRIS map. 
To do so we take advantage of the fact that each final IRIS map is 
composed of a combination of up to three HCONs. Here we argue that 
the difference between two HCONs is dominated by the instrumental 
noise and that the 
power spectrum of such a difference map is equal to the power spectrum 
of the noise, multiplied by a scaling factor that depends on the coverage 
maps.

If we make the assumption that the noise level $B$ of the IRAS detectors
is stationary, and if we neglect the photon noise (i.e. the noise level
is dominated by the readout noise),
the noise level at position $(x,y)$ of the $i^{th}$ HCON map is
\begin{equation}
\sigma_{i}(x,y) = \frac{B}{\sqrt{N_i(x,y)}}
\end{equation}
where $N_i(x,y)$ is the coverage map of the $i^{th}$ HCON map.
An example of a typical coverage map is shown on Fig.~\ref{fig_noise_iris}.

The noise level of the co-added map (HCON-0) is 
\begin{equation}
\sigma_{0}(x,y) = \frac{B}{\sqrt{N_{tot}(x,y)}}
\end{equation}
where $N_{tot}(x,y)=N_1(x,y)+N_2(x,y)+N_3(x,y)$.
Similarly the noise level of the difference between two HCON maps ($i$ and $j$) is
\begin{eqnarray}
\sigma_{i-j}(x,y) & = & \sqrt{\sigma_i^2 + \sigma_j^2}\nonumber\\
& = & B\sqrt{\frac{1}{N_i(x,y)} + \frac{1}{N_j(x,y)}}.
\end{eqnarray}
Combining the last two equations, the noise level
of the co-added map becomes:
\begin{equation}
\label{eq_noise_weight}
\sigma_0(x,y) = \sigma_{i-j}
\sqrt{\frac{N_i(x,y) N_j(x,y)}{N_{tot}(x,y)(N_i(x,y)+N_j(x,y))}}
\end{equation}

To estimate the power spectrum of the noise of a given IRIS plate 
we first compute the difference map between two HCONs. The difference map
is then normalized following Eq.~\ref{eq_noise_weight} to take into
account the spatial coverage of every HCON. An example of such
a normalized difference map is shown in Fig.~\ref{fig_noise_iris}
together with its power spectrum and the power spectrum of the
corresponding IRIS plate. At small scale (high $k$) these two power 
spectra meet perfectly which validates our processing.

It is important to point out that the difference map used here to compute
the power spectrum of the noise is by no mean a noise level map.
It does not provide a way to estimate the noise level of every pixel
in the map but it is only used to estimate the global properties
of the noise (i.e. power spectrum, average level).

The average noise level of an IRIS map can be estimated 
by taking the standard deviation of the difference map.
Typical noise level values are given in Table~\ref{table_issa}
for the four bands. Within 10\%  the IRIS noise level
is the same as the ISSA noise level. 
This is due to the fact that the power spectrum of the noise
is not flat (its slope varies from -1.5 to 0). Therefore
the standard deviation of the difference map may be dominated
by large scale structures in the noise.
In such case, integral of the ISSA and IRIS noise power spectra
on selected scales allows a better comparison.
For large scales (1-10$\rm ^o$), the noise level in the IRIS plates
is about 30$\%$ smaller than that of the ISSA plates, for the four
wavelengths. At smaller scales, as explained in \S \ref{section_properties},
we gain in sensitivity only at 60 and 100 $\mu$m.
The gain is about 15-20$\%$ for scales 5-15'.

\subsection{Effective angular resolution}

\label{section_resolution}

Due to the geometry of the IRAS detectors, the IRAS data have a higher
spatial resolution in the in-scan direction than in the cross-scan
(see Table~\ref{table_issa}). 
\cite{wheelock93} have computed the PSF on few (4 to 5) point sources
in each band. They showed that the PSF is non-circular due to
difference in the in-scan and cross-scan resolution. They conclude that
the angular resolution of the ISSA plates is between 4 and 5 arcmin
in each band.

Following \cite{miville-deschenes2002b} we used the power spectrum of ISSA
plates to estimate the effective Point Spread Function (PSF) of the ISSA.
If we make the assumption that the effective PSF is described by a Gaussian, 
the power spectrum of an image dominated by interstellar
fluctuations can be expressed by the following equation:
\begin{equation}
P(k) = \exp\left(\frac{-k^2}{2\sigma_k^2}\right)\times Ak^\beta
\end{equation}
where $Ak^\beta$ describes the cirrus emission power 
spectrum \cite[]{gautier92}. We have done such a fitting for several 
IRIS plates at the 4 wavelengths. We have selected plates with relatively
bright interstellar emission (not to be affected by noise) from which we have
removed the bright point sources. The effective resolution of the four 
IRIS bands are given in Table~\ref{table_issa}. 
We have checked that these values are compatible with Gaussian fitting
on point sources.

\begin{figure*}
\includegraphics[width=\linewidth]{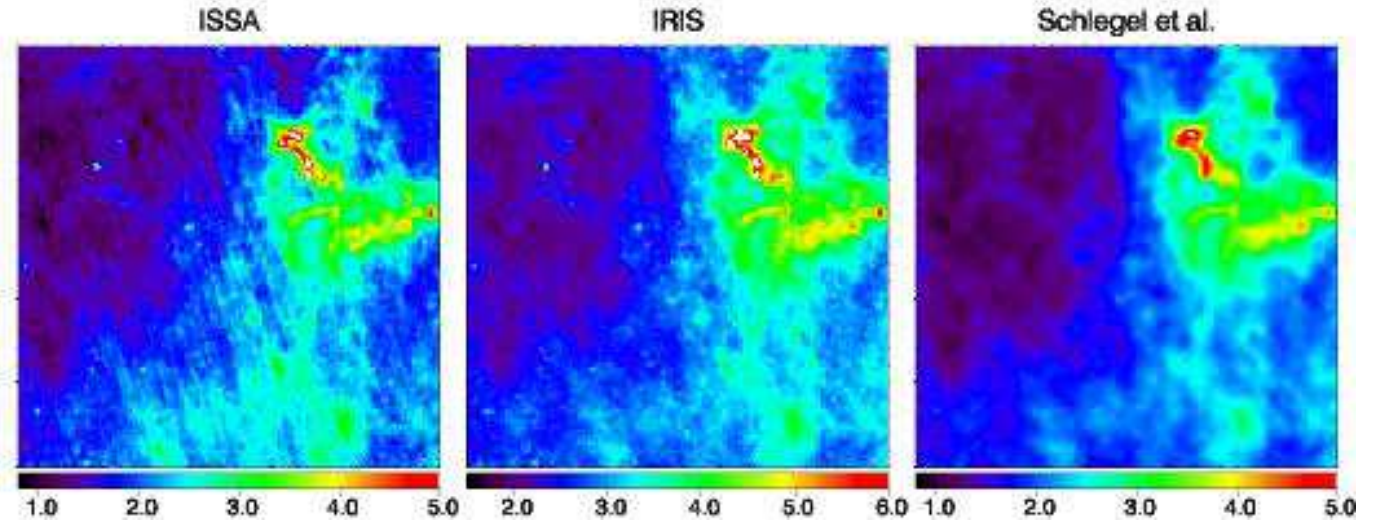}
\caption{\label{fig_compare_image} Comparison of a ISSA, IRIS and 
Schlegel et al. $250\times250$ sub-image at 100~\ump. All the maps
are in MJy/sr. The IRIS
map combined the higher angular resolution of the ISSA map
and the destriping and zero level correction 
of the Schlegel et al. map. In addition the IRIS map 
was corrected for the variation of the IRAS responsivity with 
scale and brightness and it also includes
properly calibrated point sources.}
\end{figure*}

\begin{figure}
\includegraphics[width=\linewidth]{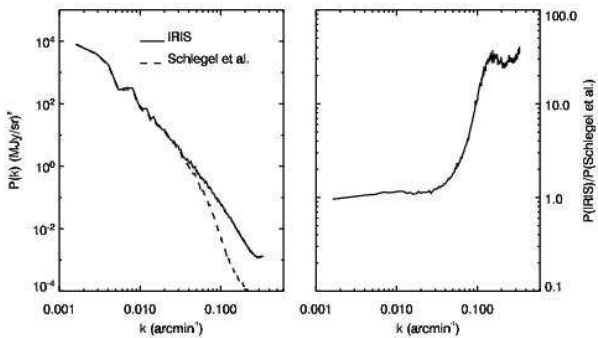}
\caption{\label{fig_compare_pk} Comparison of the power
spectrum of a typical 100~\um image with IRIS and Schlegel et al. 
{\bf Left:} Power spectra of the IRIS (point sources subtracted) and
Schlegel et al. plates. {\bf Right:} Ratio of the IRIS  
and Schlegel et al power spectra. The two plates have a very similar
power spectrum except at high-$k$ due to the higher angular
resolution of the IRIS plate.}
\end{figure}

\subsection{Calibration uncertainties}

At 60 and 100~\ump, the responsivity correction we applied at scales smaller 
than 1.25$^\circ$ depends on an interpolation between the DC/AC factor 
(responsivity at 5' scale) and the responsivity at scale of 1.25$^\circ$
(deduced from the comparison of IRAS and DIRBE with scale and brightness - 
see \S~\ref{section_gain}). 
The largest responsivity difference between these two scales is for bright 
regions where it reaches $\sim$20\% (see Fig.~\ref{fig_gain}).
We estimate that the typical uncertainty on the responsivity  
at scales smaller than 1.25$^\circ$ is $\sim$5\%, for the four bands.

At scales larger than 1.25$^\circ$ the IRAS data are forced to match
the DIRBE data by applying an offset map (see \S~\ref{section_offset}).
The application of such an offset map takes care of responsivity
variations at scales larger than 1.25$^\circ$ but it also sets the
zero level correctly.
The uncertainty on the responsivity at large scales is thus limited
by the DIRBE gain uncertainties which varies from 5.1 to 15.1\% 
(see Table~\ref{table_issa}). Considering the relatively small
uncertainty of the responsivity at sub-degree scales we estimate
that the overall uncertainty on the IRIS calibration is dominated
by the uncertainty on the DIRBE data (including the uncertainty
on the zodiacal model).\\

In the comparison of the IRAS and DIRBE data, an additional uncertainty
comes from the slight differences of the system spectral responses. To take into account
these differences, the DIRBE data should have been multiplied by the ratio of the
IRAS to DIRBE color correction prior to any comparison with the IRAS data
(see \S~\ref{section_color_correction}). This ratio 
depends on the details of the wavelength response of the system
but also on the
shape of the intrinsic energy distribution that is unkown for each sky direction (especially
at short wavelengths). 
No correction was therefore applied to the DIRBE data.
This induces an additional uncertainty  at 100, 60, 25 and 12 microns
that is less than typically 6, 10, 3 and 10$\%$ for the diffuse medium
respectively.

\subsection{Comparison with Schlegel et al.}

A first attempt to produce well calibrated 100~\um IRAS maps was done by
\cite{schlegel98}. These authors improved the destriping and 
zero level of the ISSA plates in a similar fashion as we did. 
One of the great improvement of the IRIS 100 $\rm \mu$m 
data compared to \cite{schlegel98} is that we take into account the
variation of the responsivity with brightness at scales smaller
than 1$^\circ$ (while they apply a constant factor of 0.87). 
In addition the product given by \cite{schlegel98} does not 
include point sources. 

In Fig.~\ref{fig_compare_image} we present a comparison of a 100~\um
$250\times250$ pixels sub-image of the same region in
ISSA, IRIS and \cite{schlegel98}. One striking feature of this figure
is the striping that is completely removed in the IRIS and \cite{schlegel98}
images. One 
feature of the IRIS product is also the fact that it includes point sources.
One would also note the difference in angular resolution
between the IRIS and \cite{schlegel98} data. Contrary to \cite{schlegel98}
we have taken great care to keep the full ISSA resolution. 
The difference in angular resolution between IRIS and \cite{schlegel98}
is well seen in Fig.~\ref{fig_compare_pk} where we compare the power
spectrum of a typical 100~\um map\footnote{Point sources have been
subtracted from the IRIS map prior to compute the power spectrum.}. 
There is a very good agreement between
these two data sets at scales larger than $\sim30'$ but there is
a significant difference at smaller scales due to the difference in
the responsivity correction and also to the lower resolution of the 
\cite{schlegel98} data (6.1').

\section{Conclusion}
We have built a new generation of IRAS maps that are
rescaled to an absolute calibration
coherent with both the DIRBE and the IRAS Point Source Catalog
photometry.
These maps are also much less contaminated by stripes
than the previous IRAS images.
This new generation of IRAS maps is named IRIS
for Improved Reprocessing of the IRAS Survey.
The IRIS data are available at {\bf http://www.cita.utoronto.ca/~mamd/IRIS} and
{\bf http://www.ias.u-psud.fr/iris}. They can also be obtained from {\bf http://www.ipac.caltech.edu}
and through the Centre de Donn\'ees astronomiques 
de Strasbourg (CDS - {\bf http://www.cdsweb.u-strasbg.fr}) via
{\it Aladin}. The data products are detailed in Appendix~\ref{data_products}.\\

Compared to the previous products (namely the ISSA and \cite{schlegel98}), 
we made a major improvement of the calibration 
and zero level at all wavelengths
(but especially at 12 and 25~\ump) by comparing with DIRBE 
and remove significantly the stripes using an
efficient iterative process  
that involve a wavelet decomposition of the map.
Considering the relatively small
uncertainty of the IRAS responsivity at sub-degree scales we estimate
that the overall uncertainty on the IRIS calibration is dominated
by the uncertainty on the DIRBE data (including the uncertainty
on the zodiacal model). One other great interest of the IRIS maps is that 
they contain all the IRAS small scale information (including point-like
sources) properly calibrated. 
With IRIS, we have reached the limit of the post-processing of the ISSA plates.
Improvements over the IRIS product will be only possible 
by working directly on the raw IRAS data.\\

The IRIS data will be of great use for the whole community.
In the context of the numerous ongoing and future 
infrared/submm/mm missions (Spitzer, Herschel, Planck, JWST, ASTRO-F...), 
they will be very helpful to prepare and analyse the observations.
In particular, with an angular resolution comparable to the 
high frequency bands of Planck, these data will be very complementary, 
providing unique dust templates. These data can also be used by themselves 
to conduct still original science. 
As an example, in a forthcoming paper, we will present a study of the properties 
of the cirrus dust emission.

\section*{Acknowledgments}
We warmly thanks S. Wheelock, N. Gautier, F. Boulanger
and J.-L. Puget for the help and support they give to this work.
Thanks to all the IRAS team for providing us with the great legacy
ISSA product. The processing of the IRIS plates has been done on the 
Beowulf cluster of the Canadian Institute for Theoretical Astrophysics.

\begin{appendix}
\section{Point source identification}

\label{section_point_source}

\begin{figure*}
\includegraphics[width=\linewidth]{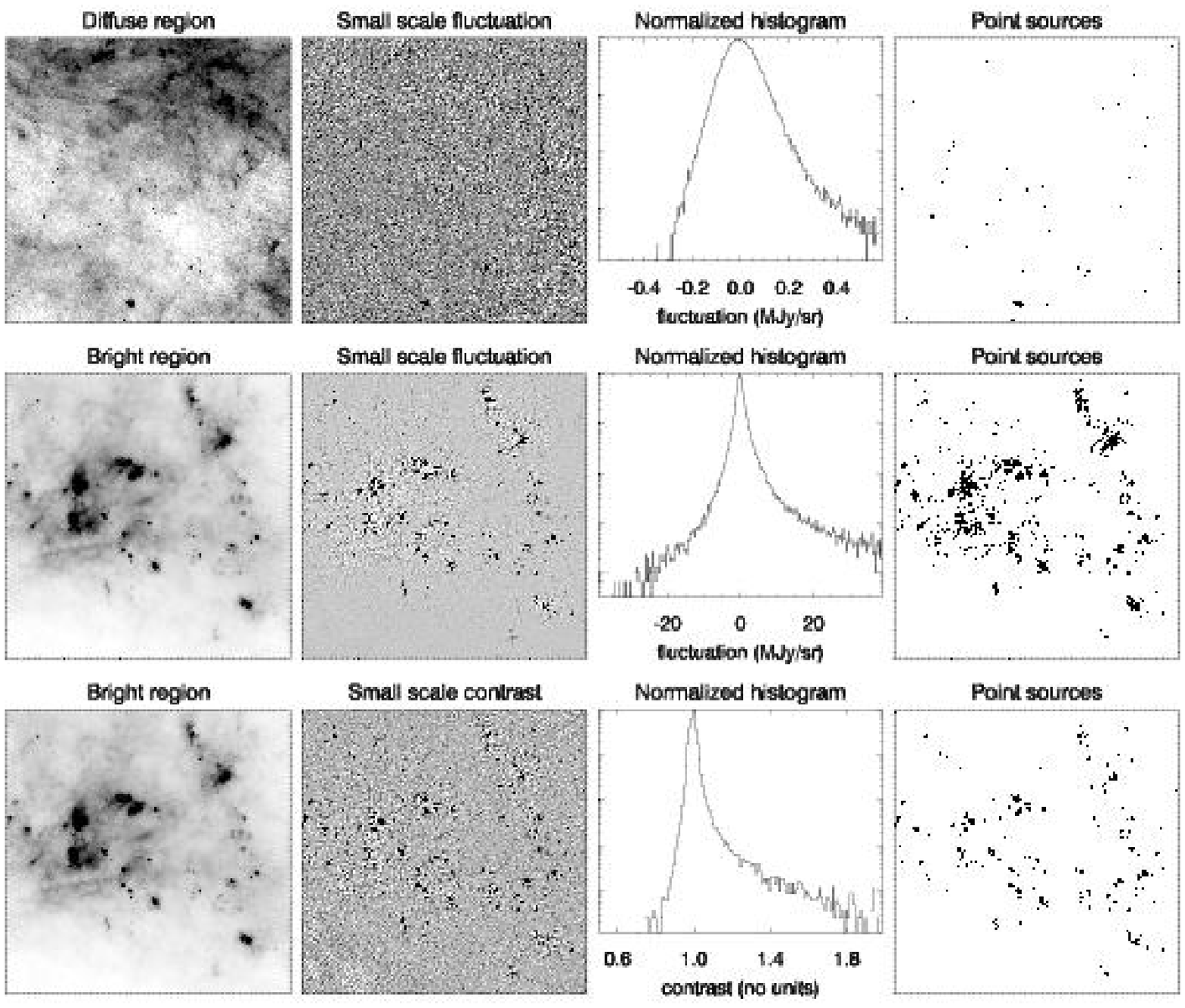}
\caption{\label{fig_clip_source} Exemple of point source identification.}
\end{figure*}

One important part of the post-processing of the ISSA maps
presented here is related to the identification and removal of point sources, 
or more generally of small scale structures.
The removal of point sources is used at several steps in the
processing. As the calibration of the diffuse emission and
of point sources are not the same in the ISSA plates, they 
have to be treated separately. Point sources 
identification is also used for the deglitching and destriping of the 
HCONs and for the determination of the zero level map.

The difficulty here is to develop a point source identification method 
that is equally efficient in very diffuse regions where the
noise is significant and in crowded Galactic plane fields. 
The poor-man method to identify point sources is to apply 
a boxcar filtering to the image and to look for highly deviant 
small scale fluctuations.
For images with relatively homogeneous background
the histogram of small scale brightness fluctuation 
is dominated by a quasi-Gaussian core, partly due to noise,
with a positive wing due to point sources 
(see Fig.~\ref{fig_clip_source}). In this case point
sources can easily be identified as beeing
all the pixels with fluctuation values higher than $n\sigma$ where
$\sigma$ is the width of the Gaussian core and $n$ is a given
threshold.

For images with strongly varying diffuse emission this method does not 
work as well. In Fig.~\ref{fig_clip_source} (middle row) we present the
identification of point sources with this traditional sigma-clipping 
method. In this case the shape of the small-scale fluctuation histogram
is very different and a much larger fraction
of the pixels in this image have a small scale fluctuations 
above $n\sigma$, where $\sigma$ is the width of the central component 
of the histogram. The result (middle-right panel) is that most of the
structure in the bright parts of the image has been identified as point 
sources.

For bright regions the small-scale fluctuations are not dominated
by noise fluctuations. The core of the small-scale fluctuation
histogram comes from the diffuse emission itself.
As shown by \cite{gautier92}, the amplitude of the diffuse
emission fluctuations $P(l)$ at a given scale $l$ increases with the
average brightness $B_0$:
\begin{equation}
\label{eq_gautier}
P(l) \propto B_0^\alpha
\end{equation}
\cite{gautier92} gave $\alpha=1.5$. 
This indicates clearly that 
the fluctuations level of the diffuse emission 
increases with brightness. This is clearly seen in Fig.~\ref{fig_clip_source}
where bright regions have generally stronger small scale
fluctuations. 

To take this effect into account we rather use the contrast
of the small scale fluctuations (i.e. fluctuations / brightness)
to identify point sources, instead of the small scale fluctuation
level itself. This way we take into account the
variation of the diffuse emission fluctuations with brightness
(see Eq.~\ref{eq_gautier}). The result of this method is shown
on the bottom row of Fig.~\ref{fig_clip_source}; 
much less pixels were clipped.
This method is much more flexible than the traditional sigma-clipping
method as it adapts itself to the background level. 

At this point it is important to mention that the contrast method
can't be applied in diffuse region where noise contributes
significantly to the small scale brightness fluctuations. 
For each band we used the traditional sigma-clipping for regions
with boxcar filtered brightness lower than 3, 3, 4 and 4 MJy/sr at 
12, 25, 60 and 100~\um respectively. For regions with brightness higher 
than those values we used the contrast method to identify point sources.

\section{IRAS and DIRBE filters}

\label{section_color_correction}

\begin{figure*}
\includegraphics[width=\linewidth]{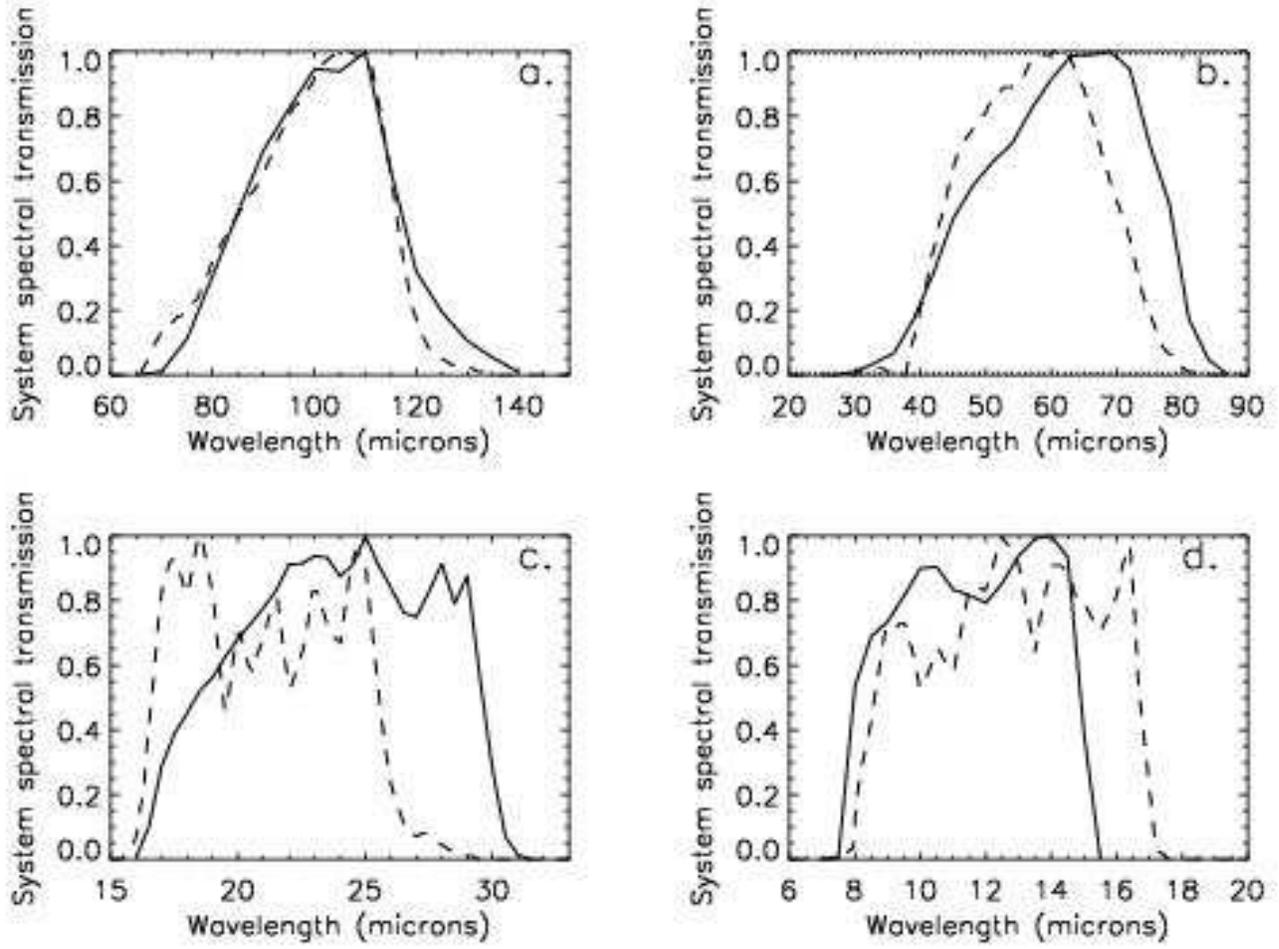}
\caption{\label{fig_filters} System spectral responses for the IRAS (continuous lines)
and DIRBE (dashed lines) 100, 60, 25 and 12 $\rm \mu$m bands (Figs. a. to d. respectively).}
\end{figure*}

Following the IRAS convention, the DIRBE spectral 
intensities I$_{\rm \nu}$ are
expressed in MJy/sr at fixed nominal frequencies assuming the source 
spectrum is $\rm \nu I_{\nu}$=constant (i.e. constant intensity
per logarithmic frequency interval).
Since the source spectrum has not a constant intensity
per logarithmic frequency interval, a color correction
has to be applied to obtain an accurate intensity. 
The color correction factor is defined such that:
\begin{equation}
\label{eq_cc}
I_{\nu_0}(actual) = I_{\nu_0}(quoted)/ cc
\end{equation}
where $I_{\nu_0}(actual)$ is the actual specific intensity
of the sky at frequency $\nu_0$, $I_{\nu_0}(quoted)$ is the corresponding
value given with the IRAS or DIRBE convention
and $\nu_0$ is the frequency corresponding to the nominal wavelength 
of the considered band. With these definitions:
\begin{equation}
cc= \frac{\int (I_{\nu}/I_{\nu_0})_{actual} R_{\nu} d \nu}{\int (\nu_0/\nu)  R_{\nu} d \nu}
\end{equation}
where $(I_{\nu}/I_{\nu_0})_{actual}$ is the actual specific intensity 
of the sky normalised to the intensity at frequency $\nu_0$ and $R_{\nu}$
is the system spectral response. The spectral responses 
are given in Fig. \ref{fig_filters} for the four IRAS and DIRBE bands.\\

\begin{table}
\begin{center}
\begin{tabular}{|c|c|c|} \hline
Wavelength & (T (K), $\beta$) & $\frac{cc_{IRAS}}{cc_{DIRBE}}$ \\ \hline
100 $\mu$m & (14, 2) & 1.1019 \\  
 & (16, 2) & 1.0737 \\ 
 & (18, 2) & 1.0525 \\ 
 & (20, 2) & 1.0354 \\ 
 & (22, 2) & 1.0211 \\
 & (14, 1.5) & 1.1146 \\
 & (16, 1.5) & 1.0852 \\
 & (22, 1.5) & 1.0321 \\ 
 & (-, -) & 1.0510  \\ \hline
60 $\mu$m  & (40, 1) &  1.0476\\ 
 & (60, 1) & 0.9452 \\ 
 & (40, 0) & 1.1187 \\ 
 & (60, 0) &0.9947  \\
 & (-, -)  & 1.1045  \\  \hline
25 $\mu$m  & (-, 0) & 0.8941 \\ 
 & (-, 1) & 0.8036  \\
 & (-, 2) & 0.7265 \\
  & (-, -) &  0.9728 \\ \hline
12 $\mu$m  & (-, 0) & 1.0868 \\ 
 & (-, 1) & 1.1723 \\
 & (-, 2) & 1.2552 \\
 & (-, -) & 1.0938  \\ \hline
\end{tabular}
\end{center}
\caption{\label{tbl:cc}Ratio of IRAS to DIRBE color 
corrections for spectra in the form 
$I_{\nu}\propto \nu^{\beta} B_{\nu}(\rm T)$. 
When no values are given for the temperature, the considered
spectrum is in the form $I_{\nu}\propto \nu^{\beta}$.
When no values
are given for both the temperature and spectral index, 
the considered spectrum
is the standard diffuse ISM spectrum for 10$^{\rm 20}$ at/cm$^{\rm 2}$.}
\end{table}

When we compare the DIRBE and IRAS intensities, we compare
$I_{\nu_0}(quoted)_{DIRBE}$ and $I_{\nu_0}(quoted)_{IRAS}$. 
For the same point on the sky and in the same beam (i.e. when IRAS
data are convolved by the DIRBE beam), 
$I_{\nu_0}(actual)_{DIRBE} =I_{\nu_0}(actual)_{IRAS}$.
This leads to:
\begin{equation}
\frac{I_{\nu_0}(quoted)_{IRAS}}{I_{\nu_0}(quoted)_{DIRBE}}=
\frac{cc_{IRAS}}{cc_{DIRBE}}
\end{equation}
{\it Therefore when comparing the DIRBE and IRAS data to calibrate the 
IRAS data, we should correct the DIRBE data by the ratio of color
corrections}. This ratio is varying across the sky. It is quite
easy to compute it in the Far-Infrared at large scales where we can infer the 
dust emissivity and temperature using DIRBE data (using an iterative procedure).
At smaller scales and at 12 and 25 $\mu$m, it is impossible to compute this ratio
since we do not know the exact shape of the intrinsic energy distribution
for most of the sky.
As an example, ratios of color corrections are given for different spectral energy
distribution in Table \ref{tbl:cc}.

\section{Data products}

\label{data_products}
The ``standard'' IRIS data (``standard'' means for the processing described in this
paper) are available at http://www.cita.utoronto.ca/~mamd/IRIS and http://www.ias.u-psud.fr/iris.
On these two sites the data are available in the original ISSA 
format: 430 maps of $500\times 500$ pixels for each band. The individual HCON
can also be downloaded from these sites.
In the near future we expect to provide Healpix version of these maps on these
two sites. We also expect to provide versions of the maps
without point sources and with an improved zodiacal light subtraction.

The IRIS data product is also available through the Centre de Donn\'ees astronomiques 
de Strasbourg (CDS - http://www.cdsweb.u-strasbg.fr) via
{\it Aladin} and at the Infrared Processing and Analysis Center
(IPAC). 

\end{appendix}

\end{document}